\documentclass[preprint,1p,number]{elsarticle}
\usepackage{dcolumn}
\usepackage{graphicx}
\usepackage{amsmath}
\usepackage{amssymb}
\biboptions{sort&compress}
\usepackage{float}
\usepackage{tabularx}
\usepackage{xcolor}
\usepackage{enumerate}
\usepackage{bm}

\begin{document}

\begin{frontmatter}
\title{A Laplace-like formula for the energy dependence \\of the nuclear level density parameter}
\author[cbu]{B.~Canbula\corref{cor1}}
\ead{bora.canbula@cbu.edu.tr}
\cortext[cor1]{Corresponding author}
\author[cbu]{R.~Bulur}
\author[cbu]{D.~Canbula}
\author[cbu]{H.~Babacan}
\address[cbu]{Department of Physics, Faculty of Arts and
Sciences,\\Celal Bayar University, 45140, Muradiye, Manisa,
Turkey}
\date{\today}
\begin{abstract}
Collective effects in the level density are not well understood, and 
including these effects as enhancement factors to the level density 
does not produce sufficiently consistent predictions of observables. 
Therefore, collective effects are investigated in the level density 
parameter instead of treating them as a final factor in the level 
density. A new Laplace-like formula is proposed for the energy 
dependence of the level density parameter, including collective 
effects. A significant improvement has been achieved in agreement 
between observed and predicted energy levels. This new model can 
also be used in both structure and reaction calculations of the 
nuclei far from stability, especially near the drip lines.
\end{abstract}
\begin{keyword}
nuclear level density \sep semi-classical \sep collective motion 
\sep harmonic oscillator \sep Coulomb potential 
\sep rotational modes \sep vibrational modes 
\end{keyword}
\end{frontmatter}

\section{Introduction}\label{introduction}

Nuclear level density (NLD), which is the number of the excited levels 
around an excitation energy, has been studied for nearly eight 
decades. The knowledge about the NLD is the key 
of the accurate Hauser-Feshbach calculations for the 
compound-nucleus cross sections. It becomes an obligation to use 
the level density function in the case of incompleteness of the 
experimental information on the energy levels, or at high 
excitation energies, which levels become very narrowly spaced, 
or even continuous. The first study on this subject was 
proposed by Bethe \cite{bethe1937}, who introduced the Fermi 
gas model, and many authors have been studied on this subject 
extensively with several methods 
\cite{gilbert1965,demetriou2001,hilaire2006,nerlopomorska2002,nerlopomorska2006a,nerlopomorska2006b,newton1956,dilg1973,holmes1976,krusche1986,edigy1988}. 
Although it is highly desired to use microscopic models, 
phenomenological models are still useful and popular due to 
their simplicity and ease of application. On the other hand, 
these models usually have several free parameters to be 
adjusted the experimental data, namely the mean resonance 
spacings and the discrete level schemes. The fitting and the 
subsequent extrapolation of the parameters are the main 
limitations to use these models reliably for the nuclei far 
from stability. After the pioneering work of Tanihata 
\cite{tanihata1985}, the nuclei near the drip lines have 
been subject of interest because of their unusual properties. 
Therefore, to propose a level density model, which can be used 
as a reliable tool for the theoretical calculations of the 
reactions involving light exotic beams, is an outstanding 
problem in nuclear structure.

The other crucial problem of the level density is the 
collective enhancement. The coherent collective nuclear 
excitations cause an increase in the level density and play 
a dominant role at the low-energy region before damping with 
the increasing excitation energy. Therefore, without taking 
into account these effects, it is impossible to describe the 
first few low-lying excited states of the nucleus. The collective 
effects can be separated into two parts, namely vibrational 
and rotational. While the rotational excitations make 
contributions to the level density only for deformed shapes, 
the vibrational effects should be considered even for the 
spherical nuclei. In spite of many studies on the collective 
effects in the level density 
\cite{bjornholm1973,hansen1983,junghans1998}, the results 
are not at the level of expectation. Hence, 
this problem still remains unsolved and requires further 
investigation.

In the light of the above discussion, the objective of 
this paper is to propose a new method to include the collective 
effects into the level density formalism and improve the usability 
of the level density in the reaction calculations of the exotic 
nuclei. This paper is organized as follows: A brief introduction 
of the nuclear level density is given in Section \ref{theory}. 
The results of the calculations are presented in Section 
\ref{results}. Finally, in Section \ref{conclusions}, we 
summarize our results and discuss their significance.

\section{Theory}\label{theory}

According to the Fermi gas model, nucleus treated as a system of 
non-interacting nucleons and collective levels are absent, therefore 
excited levels arise only from the single-particle states with 
equally spaced. Under these assumptions, the level density of a 
double fermion system, which is formed from protons and neutrons, 
is given as a function of effective energy $U=E_{x}-\Delta$, 
level density parameter $a$, spin $J$, spin cut-off parameter 
$\sigma^{2}$, with equiparity distribution
\cite{bethe1937,ericson1960}
\begin{equation}
\label{eq:leveldensity}
\rho(U,J,\Pi) = \frac{1}{2} \frac{2J+1}{2 \sqrt{2 \pi} \sigma^{3}}
\mathrm{exp} \left [ - \frac{\left ( J + \frac{1}{2} \right
)^{2}}{2 \sigma^{2}} \right ] \frac{\sqrt{\pi}}{12} \frac{\mathrm{exp} [2 \sqrt{a U}]}{a^{1/4} U^{5/4}}
\end{equation}
Spin cut-off parameter can be written as $\sigma^{2}=T I / \hbar^{2}$ 
in the simplest form where $T$ is the nuclear temperature, and $I$ is 
the moment of inertia. The energy shift 
$\Delta = \delta + n \frac{12}{\sqrt{A}}$ where $n$ is $-1$ for 
odd-odd, $1$ for even-even, $0$ for odd nuclei and $\delta$ remains 
as an adjustable parameter to fit. The total level density can be 
obtained by summing \eqref{eq:leveldensity} over all spins 
\begin{equation}
\label{eq:totalleveldensity}
\rho^{\mathrm{tot}}(U) = \frac{1}{12 \sqrt{2} \sigma} 
\frac{\mathrm{exp}[2 \sqrt{a U}]}{a^{1/4} U^{5/4}}.
\end{equation}
This equation provides a simple and successful description of 
the level density, especially around the neutron separation energies, 
but it also causes a divergence problem when excitation energy goes 
to zero. This problem remained unsolved until 1985 \cite{grossjean1985}, 
and Demetriou \cite{demetriou2001} proposed a convenient 
solution, which is also used in this study, in 2001.

The main variable of the NLD is the level density parameter $a$ 
and commonly given by Ignatyuk's \cite{ignatyuk1975} formula depending 
on the excitation energy as given below
\begin{equation}
\label{eq:ldp_ignatyuk}
a(U) = \tilde{a} \left( 1 + \delta W \frac{1 - \exp [-
\gamma U]}{U} \right).
\end{equation}
$\delta W$ is the microscopic correction term of the liquid drop 
mass formula and $\tilde{a}$, the asymptotic level density parameter, 
is the limit value of $a$ that is reached at high excitation energies, 
especially beyond the neutron separation energy. The damping parameter 
$\gamma$ is given as $\gamma=\gamma_{1}/A^{1/3}$ where $\gamma_{1}$ 
is an adjustable parameter that determines how rapidly $a$ goes to 
$\tilde{a}$ and the direction of this damping depends on the sign of the 
$\delta W$.

In the early studies of the level density, the level density parameter 
was taken to consist of only its asymptotic value, therefore, it was independent 
of the excitation energy. This parameter is usually deduced from the 
experimental data by using a parameterized equation 
\cite{koning2008,iljinov1992,bartel2006} as well as it can be 
calculated theoretically from the proton and neutron single-particle 
level densities at corresponding Fermi energies 
\begin{equation}
\label{eq:theoreticalasyp}
\tilde{a} = \frac{\pi^{2}}{6} \left [ g_{p}(E_{F}^{p}) +
g_{n}(E_{F}^{n}) \right ].
\end{equation}
One can use the semi-classical 
approximation to calculate the single-particle level density 
at a single-particle energy $\varepsilon$ with the spin degeneracy 
\cite{bracksemiclassicalphysics,salasnich2000} 
\begin{equation}
\label{eq:semiclassicalspld}
g(\varepsilon) = \frac{2}{\pi} \left( 
\frac{2 m}{\hbar^{2}} \right)^{3/2} \int r^{2} \sqrt{\varepsilon - V(r)} \, \mathrm{d}r
\end{equation}
where $V(r)$ and $m$ are the average simple potential and the mass 
of the nucleon, respectively. The value of the Fermi energy can be 
found from the following conservation condition between the nucleon 
number $\mathcal{N}_{\alpha}$ and the single-particle level 
density $g_{\alpha}$ 
\begin{equation}
\label{eq:semiclassicalconservation}
\mathcal{N}_{\alpha} = \int_{- \infty}^{E_{F}^{\alpha}} g_{\alpha}(E) dE, 
\qquad\qquad \mathcal{N}_{\alpha} = \lbrace N,Z \rbrace.
\end{equation}

\begin{table}[b]
\caption{\label{tab:previouscomparison}
Goodness-of-fit estimators of the existing phenomenological 
level density models.
}
\centering
\begin{tabularx}{\textwidth}{Xllrr}
\hline
\hline
Model&Type&$f_{\textrm{rms}}$&$f_{\textrm{lev}}$\\
\hline
BSFGM \cite{koning2008} & Effective & $1.68$ & $28.5$ \\
BSFGM \cite{koning2008} & Collective & $1.71$ & $35.3$ \\
CGCM \cite{koning2008} & Effective & $1.76$ & $24.2$ \\
CGCM \cite{koning2008} & Collective & $1.77$ & $47.8$ \\
GSM \cite{koning2008} & Effective & $1.78$ & $28.0$ \\
GSM \cite{koning2008} & Collective & $1.94$ & $47.4$ \\
\hline
\end{tabularx}
\end{table}

With the model described so far it is possible to calculate  
two observables, mean resonance spacings 
\begin{equation}
\label{eq:meanresonancespacing}
\frac{1}{D_{0}^{\mathrm{theo}}} = \sum_{J=\left| I-\frac{1}{2}
\right|}^{J=I+\frac{1}{2}} \rho(S_{n},J,\Pi)
\end{equation}
and cumulative levels up to an excitation energy 
$E_{x}$ from the lower-level $N_{L}$ with the energy $E_{L}$, 
\begin{equation}
\label{eq:cumulativelevels}
N_{\mathrm{cum}}(E_{x}) = N_{L} + \int_{E_{L}}^{E}
\rho^{\mathrm{tot}}(E_{x}) dE_{x}
\end{equation}
with the adjustable parameters $\delta$ and $\gamma_{1}$. 
For $N$ nuclei, the quality of these calculations is given by 
the rms deviation factor for mean resonance spacings 
\begin{equation}
\label{eq:frms}
f_{\mathrm{rms}} = \exp \left[ \frac{1}{N} \sum_{i=1}^{N} \left(
\ln \frac{D_{0,i}^{\mathrm{theo}}}{D_{0,i}^{\mathrm{exp}}} \right
)^{2} \right]^{1/2}
\end{equation}
and the average goodness-of-fit estimator for discrete levels 
\begin{equation}
\label{eq:flev}
f_{\mathrm{lev}} = \frac{1}{N} \sum_{i=1}^{N}
\sum_{k=N_{L}^{i}}^{N_{U}^{i}} \frac{\left[
N_{\mathrm{cum}}^{i}(E_{k}) - k \right ]^{2}}{k}
\end{equation}
from a lower-level $N_{L}$ with the energy $E_{L}$ to an 
upper-level $N_{U}$ with the energy $E_{U}$ \cite{capote2009}. 

Table \ref{tab:previouscomparison} shows the agreement between 
the experimental data and predicted observables from the 
phenomenological models, for which each model has two different types 
according to their ways of handling the low-lying collective 
levels. Effective models try to describe all excited levels, 
including the collective ones by fitting the adjustable 
parameters to experimental data. In contrast to this approach, 
collective models consider the total level density given by Eq. 
\eqref{eq:totalleveldensity} as an intrinsic level density, 
which describes only pure single particle excitations, and use 
enhancement factors that explicitly account the collective effects 
\begin{equation}
\label{eq:collectiveleveldensity}
\rho(U)=K_{\mathrm{rot}} K_{\mathrm{vib}} \rho_{\mathrm{int}}(U)
\end{equation}
where $K_{\mathrm{rot}}$ and $K_{\mathrm{vib}}$ are the 
coefficients for the rotational and vibrational enhancement 
respectively. Even if it seems more physical to use collective 
models for theoretical calculations, the goodness-of-fit 
estimators given in Table \ref{tab:previouscomparison} show 
that significantly better agreement with experimental data 
is achieved when effective models are used. 

\begin{figure}[htb]
\begin{center}
\includegraphics{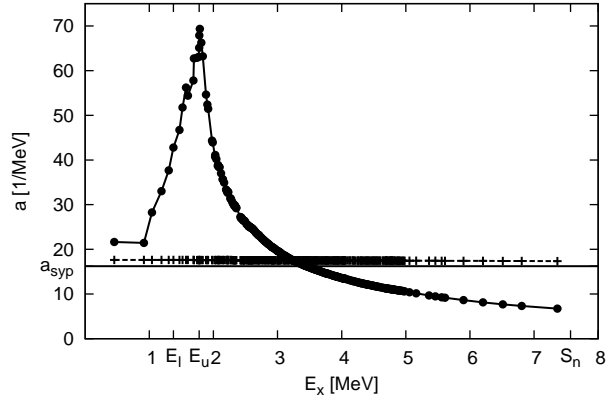}
\end{center}
\caption{\label{fig:adjustedldp} The results of level density 
parameter fit for ${}^{146}\mathrm{Nd}$. Fitted values (close 
circles) are compared to calculated values (plus signs) from 
Ignatyuk's formula \eqref{eq:ldp_ignatyuk} and calculated 
asymptotic level density parameter \eqref{eq:theoreticalasyp} 
(solid line). }
\end{figure}

This situation motivated us to search for a different method that 
can deal with collective effects and produce more accurate 
predictions than existing phenomenological models. To use 
enhancement factors for the level density expression is an 
obviously deficient and delayed attempt for describing the 
collective levels, therefore these effects must be included 
to level density calculations from the beginning. The most 
suitable candidate to include these effects seems to be 
the level density parameter. Since we know very little about 
the collective effects, it will be reasonable to start with 
considering the level density parameter as a single free 
parameter in the model and fit it to experimental data. 
The obtained results for ${}^{146}\mathrm{Nd}$ is shown 
in Figure \ref{fig:adjustedldp}. Here ${}^{146}\mathrm{Nd}$ 
is randomly chosen because most of the nuclei exhibit 
similar behavior.

Two different striking results can be deduced from Figure 
\ref{fig:adjustedldp}. First, with the increasing excitation 
energy, the level density parameter goes to a different limit 
from the asymptotic level density parameter given by Eq. 
\eqref{eq:theoreticalasyp}. The main reason of this difference 
is that the single-particle levels bunch together around the 
Fermi energy \cite{myers1966} which we have neglected in our 
calculations. This difference corresponds to the microscopic 
(or the shell) correction energy, so we must fix the calculated 
value of the Fermi energy from Eq. 
\eqref{eq:semiclassicalconservation}. 
We may also consider a further modification to the Fermi 
energy for the pairing effects. With this modification, 
all excitation energies have been corrected by an amount 
of $\Delta$ and therefore, all excitation energies have 
been transformed into effective excitation energies. With 
these modifications, the corrected Fermi energy is given by 
\begin{equation}
\label{eq:correctedfermienergy}
E_{F}^{*}=E_{F}+S(N,Z)-\Delta
\end{equation}
where $S(N,Z)$ is the shell correction energy of the liquid 
drop model \cite{myers1966}. We did not need to use the 
adjustable parameter $\delta$, so the pairing correction is 
just given as $\Delta=n \frac{12}{\sqrt{A}}$ with $n$ is $-1$ 
for odd-odd, $1$ for even-even, $0$ for odd nuclei. This way 
of handling the pairing correction also enables to make 
calculations with the excitation energies $E_{x} \leq \Delta$. 
Consequently, the corrected asymptotic level density parameter 
is given by the equation below:
\begin{equation}
\label{eq:correctedasyp}
\tilde{a} = \frac{\pi^{2}}{6} \left [ g_{p}({E_{F}^{p}}^{*}) +
g_{n}({E_{F}^{n}}^{*}) \right ].
\end{equation}
For the excitation energies higher than the neutron separation 
energy, the level density parameter is approximately equal to the 
asymptotic level density parameter. Thus, the correct description 
of the asymptotic level density parameter is the only way to 
improve the accuracy of the model in this region.

The second and even more important result coming out from 
Figure \ref{fig:adjustedldp} is the peak around $2\,\mathrm{MeV}$. 
It is well known that the origin of this extraordinary behavior 
at this energy is collective motion of the nucleons, 
in other words; this energy corresponds to the first phonon 
state arises from the vibrational motion 
\cite{kraneintroductorynuclearphysics,rowenuclearcollectivemotion}. 
Furthermore, the energy dependence of the level density parameter 
appears as a Laplacian distribution rather than the exponential 
decay given by Ignatyuk's formula \eqref{eq:ldp_ignatyuk}. All 
these unexpected results clearly show that it is an 
appropriate method to search the collective effects in 
the level density parameter. Therefore, we propose a new 
formula for the level density parameter 
\begin{equation}
\label{eq:ldp_canbula}
a(U) = \tilde{a} \left ( 1 + A_{c} \frac{S_{n}}{U} \frac{\exp (- | U - E_{0} | / {\sigma\prime}_{c}^{3})}{{\sigma\prime}_{c}^{3}} \right ).
\end{equation}
In above equation, the excitation energy is denoted by $U$, 
but it has same value as $E_{x}$ because using Eq. 
\eqref{eq:correctedasyp} all excitation energies are 
transformed into effective ones, so this procedure may be 
called as the indirect back-shifting. The location 
of the peak can be described by the excitation energy 
of the first $2^{+}$ state of even-even nuclei and 
approximated as 
\cite{rowenuclearcollectivemotion,siegbahnalphabetagammarayspectroscopy,bohrnuclearstructure1}
\begin{equation}
\label{eq:peaklocation}
E_{0}=0.2\hbar\omega
\end{equation}
where $\hbar\omega=41/A^{1/3}\,\mathrm{MeV}$. The Laplace 
distribution is desired to spread over an excitation energy 
range from the ground state to neutron separation energy 
at least and then both shell and collective effects should 
damp with the higher energies. Therefore, the scale parameter 
of the Laplace distribution must be related to neutron 
separation energy. To achieve this aim, we define a critical 
nuclear temperature as 
\begin{equation}
\label{eq:criticaltemperature}
T_{c}=\sqrt{\frac{S_{n}}{\tilde{a}}}
\end{equation}
and we use the corresponding spin cut-off parameter at this 
temperature as the scale parameter
\begin{equation}
\label{eq:criticalspincutoff}
\sigma_{c}^{2}=\frac{T_{c}I}{\hbar^{2}}.
\end{equation}
to obtain the scale parameter ${\sigma\prime}_{c}^{3}=\sigma_{c}^{3}/\tilde{a}$. 
Here we use the perpendicular moment of inertia instead of 
the spherical one $I_{0}=0.4 M R^{2}$, and it is given 
in terms of deformation parameters \cite{hagelund1977} 
\begin{equation}
\label{eq:deformedmomentofinertia}
I=I_{0} \left [ 1 + \sqrt{\frac{5\pi}{16}}\beta_{2} 
+ \frac{45}{28\pi}\beta_{2}^{2} 
+ \frac{15}{7\sqrt{5}\pi}\beta_{2}\beta_{4} \right ]
\end{equation}
and provides rotational enhancement for deformed nuclei. 

The last ingredient of Eq. \eqref{eq:ldp_canbula} is 
$A_{c}$. We define this parameter as the collective 
amplitude, and it is closely related to the shell structure 
and the surface oscillations just like the low-frequency 
collective modes. Therefore, $A_{c}$ should include the 
shell correction energy, but as the scale (spin 
cut-off) parameter, it must be at the same critical 
temperature \eqref{eq:criticaltemperature}. 
The temperature dependence of the shell correction energy 
is given by 
\cite{bracksemiclassicalphysics,bohrnuclearstructure1,nerlopomorska2005} 
\begin{equation}
\label{eq:temperaturedependentsc}
S(N,Z,T)=S(N,Z)\frac{\tau}{\sinh \tau}
\end{equation}
where $\tau=2\pi^{2}T/\hbar\omega$. It would be useful to remind 
that we denote the shell correction with $S(N,Z)$, which equals to $\delta W$ in Eq. \eqref{eq:ldp_ignatyuk}, to establish the notation. 
It is correct but insufficient to use the shell correction energy 
as collective amplitude. It is crucial to take into account the 
surface oscillations to describe the collective excitations. The 
shape dependent shell correction energy can be written as 
\cite{myers1966}
\begin{equation}
\label{eq:shapedependentsc}
S(N,Z,\mathrm{Shape})=M_{\mathrm{exp}}-M_{\mathrm{LDM}}
\end{equation}
where $M_{\mathrm{LDM}}$ is the mass, which takes into 
account the small spheroidal distortions with the equation 
below: 
\begin{equation}
\label{eq:liquiddropmass}
M_{\mathrm{LDM}}=M_{0}+E\theta^{2}.
\end{equation}
$E$ is a coefficient related to the fissility parameter $x$ 
as $E=(2/5)c_{2}A^{2/3}(1-x)\alpha_{0}^{2}$ where 
$\alpha_{0}^{2}=5(a/r_{0})^{2}A^{-2/3}$. 
$\theta=\alpha/\alpha_{0}$ is the deformation magnitude in 
terms of the deformation variable $\beta^{2}$ where 
$\alpha^{2}=(5/4\pi)\beta^{2}$. For further details see Ref. 
\cite{myers1966}. $M_{0}$ is the mass of the corresponding 
spherical nucleus and defined by the well-known formula of 
the finite-range liquid-drop model 
\cite{myers1966} 
\begin{equation}
\label{eq:sphericalmass}
M_{0}=M_{N}N+M_{H}Z+E_{V}+E_{S}+E_{C}\pm \frac{11}{\sqrt{A}}
\end{equation}
where the volume energy $E_{V}=-c_{1}A$, the surface energy 
$E_{S}=c_{2}A^{2/3}$, the Coulomb energy 
$E_{C}=c_{3}\frac{Z^{2}}{A^{1/3}}-c_{4}\frac{Z^2}{A}$, the last 
term is negative for odd-odd, positive for even-even, and equals 
to zero for odd nuclei. Finally, the collective amplitude $A_{c}$ 
is defined as the shape dependent shell correction energy at the 
critical temperature 
\begin{eqnarray}
\label{eq:collectiveamplitude}
A_{c} & = & S(N,Z,T_{c},\mathrm{Shape}) \\ \nonumber
      & = & \left [ M_{\mathrm{exp}}-M_{\mathrm{LDM}} \right ] \frac{\tau_{c}}{\sinh \tau_{c}} \\ \nonumber
	  & = & \left [ M_{\mathrm{exp}}-(M_{0}+E\theta^{2}) \right ] \frac{\tau_{c}}{\sinh \tau_{c}}
\end{eqnarray}
where $\tau_{c}=2\pi^{2}T_{c}/\hbar\omega$.

\section{Results and Discussion}\label{results}

With the level density model described so far, both global 
and local calculations can be made. In the global calculation, 
the asymptotic level density parameter must be obtained 
analytically by using Eq. \eqref{eq:correctedasyp} with the global 
potential parameters. In the present paper, we define $V(r)$ as 
the sum of central, harmonic oscillator and the Coulomb 
potential terms: 
\begin{equation}
\label{eq:potential}
V(r)=\frac{\hbar^{2}}{2mr^{2}}l(l+1)+V_{\mathrm{HO}}(r)+V_{C}(r).
\end{equation}
Harmonic oscillator potential is given by 
\begin{equation}
\label{eq:hopotential}
V_{\mathrm{HO}}(r) = \frac{1}{2} m \omega^{2} r^{2}
\end{equation}
where $\hbar\omega=41/A^{1/3}\,\mathrm{MeV}$. The coulomb 
potential of the uniformly charged sphere is 
\begin{equation}
\label{eq:coulombpotential}
V_{C}(r) = \left\lbrace
\begin{array}{ll}
\displaystyle{\frac{Z \mathrm{e}^{2}}{2 R_{C}} \left ( 3 - 
\frac{r^{2}}{{R_{C}}^{2}} \right )} & r \le R_{C} \\
\displaystyle{\frac{Z \mathrm{e}^{2}}{r}} &  r \ge R_{C}
\end{array}
\right.
\end{equation}
and charge radius $R_{C}$ is given by a simple formula 
$R_{C}=1.169\,A^{0.291}$ which is obtained from a recent 
fit \cite{bayram2013} to the latest nuclear charge 
radii data \cite{angeli2013}. In the local calculation, 
the asymptotic level density parameter is adjusted 
to the experimental data for each nucleus separately. 

\begin{figure}[htb]
\begin{center}
\includegraphics{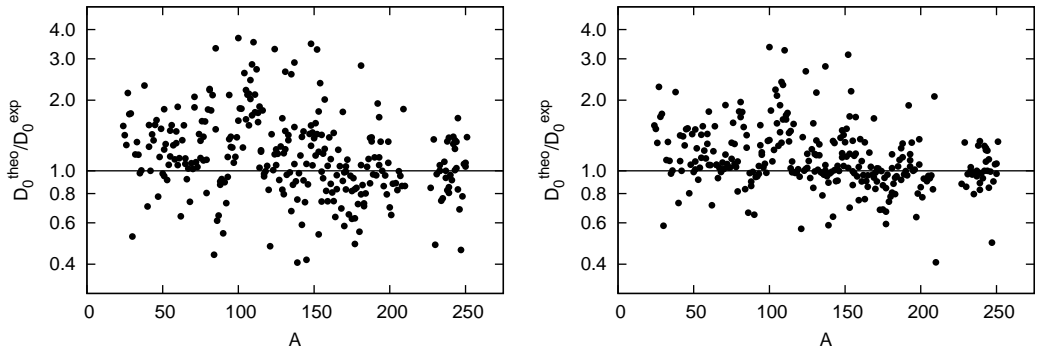}
\end{center}
\caption{\label{fig:figure2} Ratio of the predicted mean 
resonance spacing to observed value as a function of mass 
number for 289 stable nuclei. The results of the global 
and local calculations are illustrated in left and right 
panels, respectively.}
\end{figure}

Obtaining the asymptotic level density parameter by Eq. 
\eqref{eq:correctedasyp}, including the shell and the pairing 
corrections, which is one of the novelties of this paper, has 
a considerable importance in view of the wide energy range 
above the neutron separation energy. For reaction calculations 
in this regime, the definition of the asymptotic level density 
parameter is almost the only way to improve the level density 
description. The ratio of the predicted mean resonance spacings 
from the global and local calculations to the experimental data 
are plotted in Figure \ref{fig:figure2}. This ratio is the only 
indicator of the success of the level density models in the 
neutron separation energy regime. This ratio is in the range 
from $0.8$ to $2$ for the most of the nuclei, and this result 
is highly satisfactory compared to other phenomenological 
level density models. More importantly, there is no 
significance difference between the results of the global and 
local calculations, except from the general improvement in 
agreement between experiment and theory. This result is 
very promising for the reliable extrapolation of the global 
parameters for the nuclei far from stability.

\begin{figure}[htb]
\begin{center}
\includegraphics{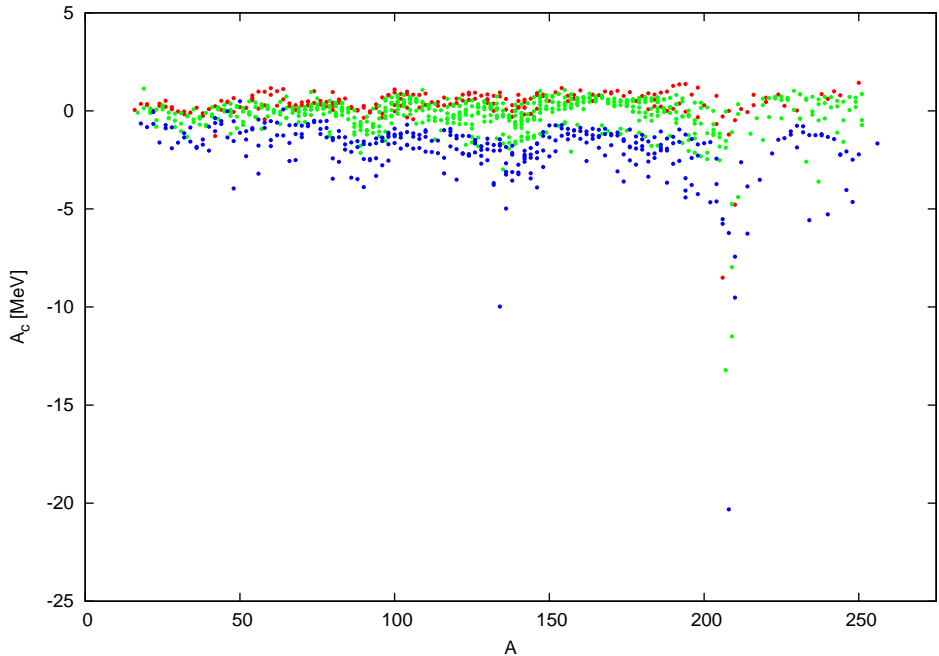}
\end{center}
\caption{\label{fig:figure3} The collective 
amplitude values obtained from the global calculation for 1136 
nuclei. The red, green and blue dots stand for the odd-odd, odd and 
even-even nuclei respectively.}
\end{figure}

The collective amplitude $A_{c}$ values obtained from 
the global calculation for 1136 nuclei, which have sufficient 
information on the discrete energy level scheme, by 
Eq. \eqref{eq:collectiveamplitude} 
are presented in Figure \ref{fig:figure3}. The newly proposed 
formula \eqref{eq:ldp_canbula} includes 
both vibrational and rotational effects. The moment of 
inertia, which is given by \eqref{eq:deformedmomentofinertia}, 
provides to include the rotational effects into the spin 
cut-off parameter. The vibrational effects are taken into 
account via both $A_{c}$ and $E_{0}$, with the shape 
dependent mass formula and the energy of the first phonon 
state, respectively. Besides the increasing magnitude of 
the collective amplitude with the mass number, its values 
are also separated for odd-odd, odd and even-even nuclei 
as clearly seen from Figure \ref{fig:figure3}. Therefore, it can 
be concluded that the pairing of the valence nucleons has 
a strong influence on the collective excitations. 

\begin{table}[htb]
\caption{\label{tab:results}
The comparison of the predictive power of the phenomenological 
level density models, including the model presented in this paper. 
The $f_{\textrm{rms}}$ covers 289 nuclei, which naturally exist on 
earth, and $f_{\textrm{lev}}$ covers 1136 nuclei, which have 
sufficient experimental information on their discrete level scheme. 
}
\centering
\begin{tabularx}{\textwidth}{Xllrr}
\hline
\hline
Model&Type&$f_{\textrm{rms}}$&$f_{\textrm{lev}}$\\
\hline
This work (Local) & Collective & $1.34$ & $0.98$ \\
This work (Global) & Collective & $1.53$ & $1.32$ \\
Semi-classical BSFGM \cite{canbula2011} & Effective & $1.12$ & $43.9$ \\
BSFGM \cite{koning2008} & Effective & $1.68$ & $28.5$ \\
BSFGM \cite{koning2008} & Collective & $1.71$ & $35.3$ \\
CGCM \cite{koning2008} & Effective & $1.76$ & $24.2$ \\
CGCM \cite{koning2008} & Collective & $1.77$ & $47.8$ \\
GSM \cite{koning2008} & Effective & $1.78$ & $28.0$ \\
GSM \cite{koning2008} & Collective & $1.94$ & $47.4$ \\
\hline
\end{tabularx}
\end{table}

The goodness-of-fit estimators, $f_{\mathrm{rms}}$ and 
$f_{\mathrm{lev}}$, values obtained from the both local 
and global calculations are given in Table \ref{tab:results}. 
All the other models, except from the model presented in this 
paper, have several adjustable parameters to be determined 
from the both mean resonance spacings and discrete level 
schemes. However, even if the reliable information on the 
discrete levels is available for over 1000 nuclei, the mean 
resonance spacings data are found only for less than 300 
nuclei, which exist naturally, the dependence to these data 
reduces the applicability and reliability of the models. 
Because of the lack of experimental data, the studies of the 
nuclei far from stability are based on an extrapolation. Our 
model does not include any of these adjustable parameters. It 
only depends strongly on the shape dependent mass formula, and 
experimental mass data are available for almost every nucleus. 
Aside from all these advantages, our model gives the best 
agreement with the experiments when compared to other models. 

\begin{figure}[htb]
\begin{center}
\begin{tabular}{ccc}
\includegraphics{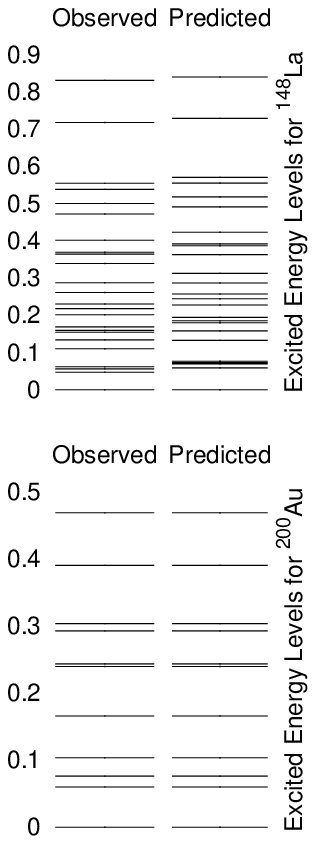} & \includegraphics{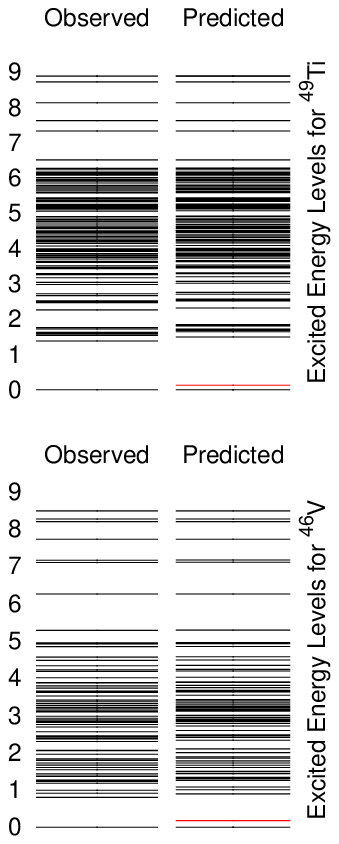} & \includegraphics{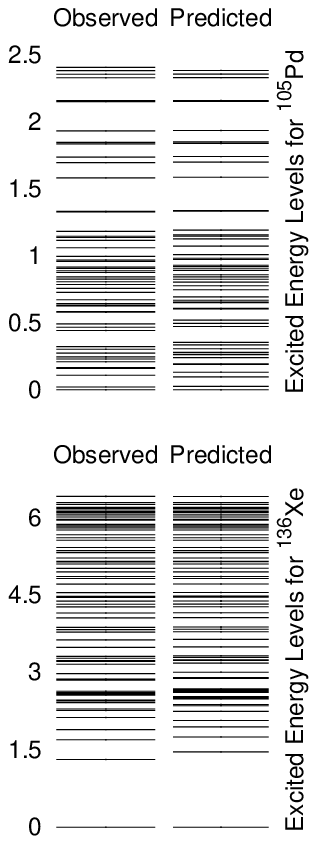}
\end{tabular}
\end{center}
\caption{\label{fig:figure4} Experimental and predicted level 
schemes of $^{148}\mathrm{La}$, $^{200}\mathrm{Au}$, 
$^{49}\mathrm{Ti}$, $^{46}\mathrm{V}$, $^{105}\mathrm{Pd}$, 
and $^{136}\mathrm{Xe}$. Predicted level schemes are obtained 
from the global calculation.}
\end{figure}

The level schemes can be constructed by using the excitation 
energies that the integral \eqref{eq:cumulativelevels} gives 
discrete integer values when these energies are used as the 
upper limit. The predicted level schemes for arbitrarily chosen 
nuclei, $^{148}\mathrm{La}$, $^{200}\mathrm{Au}$, $^{49}\mathrm{Ti}$, 
$^{46}\mathrm{V}$, and also $^{105}\mathrm{Pd}$, $^{136}\mathrm{Xe}$, 
which are known as typical collective 
nuclei, are shown in Figure \ref{fig:figure4}. 
All the predicted overlaps and gaps of energy levels are completely 
in agreement with the observed data. It is also seen from the 
right panel of Figure \ref{fig:figure4} that our cumulative level 
calculations for $^{49}\mathrm{Ti}$, $^{46}\mathrm{V}$ 
estimate a first excited state at very low energy, which is 
absent in the experimental data. Since this state is very close 
to ground state, maybe it will have a facilitating effect on 
the unresolved quasi-elastic cross-section issue of the light 
exotic nuclei \cite{keeley2009}. 

\section{Conclusions}\label{conclusions}

The analytic calculation of the asymptotic level density parameter 
including the shell and pairing effects leads the semi-classical 
approach previously described in Ref. \cite{canbula2011} to more 
physical point. In addition, the slight difference between the results 
of the global and local fit increases the reliability of the extrapolation 
of the global parameters to mass region from the stability valley to 
drip lines.

Another point which must be stressed here concerns the predicted 
levels of the excited states very close to ground state. Our model 
estimates this kind of levels, which has not observed experimentally 
yet, but they may have considerable effect in quasi-elastic cross-section 
of the light exotic nuclei. Further calculations of the quasi-elastic 
cross-section, which include these levels as inelastic absorption, 
will be necessary to understand the presence of these levels.

In summary, a new formula is proposed for the energy dependence 
of the level density parameter including collective effects. 
The results obtained by using this new formula provide an evidence 
that the level density parameter is the correct variable to include 
the collective effects and also show great improvement in agreement 
between observed and predicted energy levels as seen from Table 
\ref{tab:results}.

\section*{Acknowledgements}

This work was supported by the Turkish Science and Research 
Council (T\"{U}B\.{I}TAK) under Grant No. 112T566. Bora Canbula 
acknowledges the support through T\"{U}B\.{I}TAK PhD Program 
fellowship B\.{I}DEB-2211 Grant.

\end{document}